\documentclass[journal, a4paper]{IEEEtran}

\usepackage{listings}

\ifCLASSOPTIONcompsoc
  \usepackage[nocompress]{cite}
\else
  \usepackage{cite}
\fi

\ifCLASSINFOpdf
\else
\fi
\usepackage{array}

\usepackage{hyperref}

\begin{document}
\bstctlcite{IEEEexample:BSTcontrol}

%
\title{Decentralised Random Number Generation}

\author{
    \IEEEauthorblockN{Peter Robinson}
    \IEEEauthorblockA{PegaSys, ConsenSys and University of Queensland\\
    peter.robinson@consensys.net peter.robinson@uqconnect.edu.au}
}


\maketitle

\begin{abstract}
Decentralised random number generation algorithms suffer from the Last Actor Problem, in which the last participant to reveal their share can manipulate the generated random value by withholding their share. This paper proposes an encrypted share threshold scheme which prevents this attack.
\end{abstract}


%
\IEEEpeerreviewmaketitle

\section{Introduction}
Historically, decentralised random number generation algorithms have used a commit-reveal process in which participants submit a commitment to their randomly generated share. Once all participants have submitted commitments, each participant reveals their share. The shares are combined using a deterministic algorithm to produce the generated random value. The last participant to reveal their share, known as the Last Actor, can view all of the other shares. They can determine the impact of revealing their share, and thus decide to reveal their share or withhold their share, thus influencing the generated random value. 

DFINITY defined a Decentralised Random Beacon \cite{dfinity-consensus} using a threshold scheme and BLS Signatures to prevent the Last Actor Problem. Inspired by their work, the scheme described in this paper uses the Shamir Threshold Scheme \cite{shamir1979} combined with modulo addition to provide a decentralised random number generation scheme. This scheme is similar in some aspects to the scheme recently proposed by Drake \cite{drake2018}.

\section{Background}
In Shamir's Threshold Scheme random coefficients are generated for an equation of the form shown below, where the value $a_0$ is the secret value.

\[ Y(x) = a_0 + a_1.X + a_2.X^2 + ... + a_{m-1}.X^{m-1}  mod P \]

$n$ shares are generated. Any $m$ shares can be used to calculate the $y$ value for any $x$ value. As such, the $a_0$ secret value can be constructed from any $m$ shares.

\section{Algorithm}
\underline{Set-up}: Deploy a smart contract to the blockchain to manage the random number generation process.

\underline{Registration}: The $x$ value for each participant is the participant's Ethereum address $mod P$. Each participant generates an ephemeral RSA or ECC encryption key pair. To register, they publish their public key to the contract. The act of publishing their public key publicises their Ethereum address and hence their $x$ value. 

\underline{Calculate Random Coefficients}: All participants generate $m-1$ random coefficients for an equation in the range $1$ to $P-1$. They calculate the $y$ values for each of the participant $x$ values.

\underline{Post Commitment}: All participants post to the contract the message digest of the $y$ values for each of the participant $x$ values. Any participant which does not post commitment values drops out of the random number generation process and is fined.

\underline{Post Encrypted Y values}
All participants post to the contract the encrypted $y$ values for each of the participant $x$ values, encrypted against the public keys of each other participant. Any participant which does not post all of the encrypted $y$ values drops out of the random number generation process and is fined.

\underline{Post Private Keys and Calculate Random}
All participants post their private decryption keys. The contract then has enough information to calculate the random value and check for correctness. Correctness can be checked for by decrypting the encrypted $y$ values, checking commitments, and checking that the order of the curve that each entity posted is $m-1$. The random value is calculated as the sum of the $a_0$ values $mod P$.

To save gas, all participants post the plain text values for all of the encrypted $y$ values. All participant can off-chain check the decryptions, commitments, and order of equations. If an incorrect value is detected, this could be indicated by a call to the contract, with the contract verifying the bad value and fining the participant.  

\section{Properties}
Using commitments and asymmetrically encrypting the $y$ values means that individual attackers have to commit to a single value and can not control the release of the information. Each participant holds their own private key and publishes it once all of the encrypted $y$ values are posted, thus releasing the information for all parties to see. 

\section{Attacks}
If $m$ attackers collude they can decrypt the encrypted $y$ values as they are posted. The $m$ attackers could wait for the other $n-m$ sets of encrypted $y$ values to be posted, and then choose one or more attackers to withhold their private key, thus affecting the generated random value. The random generation process could be stopped by $n-m+1$ attackers not publishing their private keys. Doing this would mean that at least $m$ sets of $y$ values can not decrypted, and hence the $a_0$ values can not be interpolated. Both of these attacks can be countered by fining participants who do not obey the algorithm.





\bibliographystyle{IEEEtran}
\bibliography{IEEEabrv,ref}

%
%
%


\end{document}